\documentstyle[epsf,12pt]{article}

\title{\bf A time series model of CDS sequences in complete genome
}
\author{Zu-Guo Yu$^{1,2}$ and Bin Wang$^{2}$\\
   {\small $^1$Department of Mathematics, Xiangtan University, Hunan 411105,  China.
   \thanks{This is the permanent corresponding address of the first author, 
   e-mail: yuzg@hotmail.com}}\\
   {\small $^{2}$Institute of Theoretical Physics, Academia Sinica},\\
     {\small P.O. Box 2735, Beijing 100080, China. }
   }
\newcommand{\be}{\begin{equation}}
\newcommand{\ee}{\end{equation}}
\date{}
\begin{document}
\maketitle
\begin{abstract}
  A time series model of CDS sequences in complete genome is proposed. 
  A map of DNA sequence to
  integer sequence is given. The correlation dimensions and 
  Hurst exponents of CDS sequences in complete genome of bacteria
 are calculated. 
 Using the average of correlation dimensions, some interesting results are obtained.   
\end{abstract}
\vskip 0.2cm

%\pacs{ 87.10 +e, 47.53+n}


\section{Introduction}
 \ \ In the past decade or so there has been a ground swell of interest in
unraveling the mysteries of DNA. With improving of the technique of gene
clone and sequences determined, the DNA sequence data base become huge rapidly.
Doing DNA sequence analysis only use the experimental method does not fit this
rapid. Hence it becomes very important to improve new theoretical methods.
One approach that has, in just a few years, proven to be particularly fruitful
in this regard is statistical analysis of DNA sequences$^{[1-9]}$ using
modern statistical measures, including the works on the correlation properties of coding
and noncoding DNA sequences. The second approach is linguistic approach. DNA
sequence can be seen as analogous  at a number of levels to mechanisms of 
processing other kinds of languages, such as natural languages and computer
languages$^{\cite{searls}}$.  Third, using nonlinear scales method, such as fractal
dimension$^{\cite{liao,luo2,yhxc99}}$, complexity$^{\cite{YC,shen}}$. 
However, DNA sequences are more
complicated than these types of analysis can describes. Therefore, it is crucial
 to develop new tools for analysis with a view toward uncovering mechanisms
 used to code other types of information.  
 
 Since the
first complete genome of a free-living
bacterium {\it Mycoplasma genitalium} was sequenced in 1995$^{\cite{Fraser}}$,
 an ever-growing
number of complete genomes has been deposited in public databases.
The availability of complete genomes opens the possibility to
ask some global questions on these sequences. Our group also discussed the 
avoided and under-represented strings in some bacterial
complete genomes$^{\cite{yhxc99,hlz98,hxyc99}}$. In this paper, we propose a new
 model to DNA sequences, i.e. the time series model. First we want to compute the
 correlation dimension and Hurst exponents of each CDS sequence in the complete genome,
 then consider the distribution of these two quantities on complete genomes of
 Bacteria.
It is a global problem. Last we want to discuss the classification  problem
 of Bacteria using our results.
 
 \ \ For the present purpose, a DNA sequence may be regard as a sequence over
 the alphabet $\{A,C,G,T\}$ representing the four bases from which DNA is 
 assembled, namely adenine, cytosine, guanine, and thymine. For a DNA sequence,
 we define a map $f$ as following:
\begin{equation}
 f:\ \qquad \begin{array}{l} A\ \longrightarrow \ -2\\
 C\ \longrightarrow \ -1\\
 G\ \longrightarrow \ 1\\
 T\ \longrightarrow \ 2. \end{array}
 \end{equation}
Then we obtain an data sequence $\{x_k:\ k=1,2,\cdots,N\}$, where $x_k\in \{-2,-1,1,2\}$.
We formal view this sequence as a time series. 
According to the definition of $f$, the four bases $\{A,C,G,T\}$ are mapped to
four distinct value. One can also use $\{-2,-1,1,2\}$ to replace $\{A,G,C,T\}$ or
other orders of $A,G,C,T$. our main aim is distinguish $A$ and $G$ from purine,
$C$ and $T$ from pyrimidine. We expect it to reveal more information than
one dimensional DNA walk$^{\cite{peng}}$.

\section{Correlation dimension and Hurst exponent}
\begin{figure}
\centerline{\epsfxsize=12cm \epsfbox{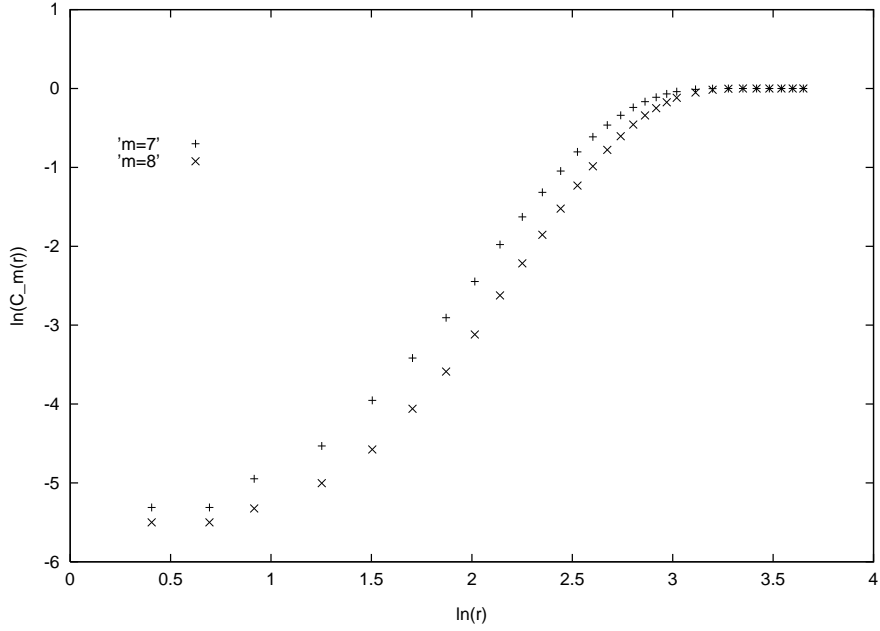}}
\caption{$\ln r-\ln C_m(r)$ figure of phage 5'UTR sequence when m=7,8.}
\label{correlfig}
\end{figure}

\ \ The notion of correlation dimension, introduced by Grassberger and
 Procaccia$^{\cite{GP1,GP2}}$,
suits well experimental situations, when only a single time series is available, it is now
being used widely in many branches of physical science.
Given a sequence of data from a computer or laboratory experiment
\begin{equation}
x_1,\ x_2,\ x_3,\cdots,\ x_N
\end{equation}
where $N$ is a big enough number. These number are usually sampled at an equal time interval
$\Delta\tau$. We embed the time series into ${\bf R}^m$, choose a time delay
 $\tau=p\Delta\tau$,
then obtain
\begin{equation}
{\bf y}_i=(x_i,x_{i+p},x_{i+2p},\cdots,x_{i+(m-1)p}),\ \ i=1,2,\cdots,N_m
\end{equation}
where
\begin{equation}
N_m=N-(m-1)p.
\end{equation}
In this way we get $N_m$ vectors of embedding space ${\bf R}^m$.

\ \ For any ${\bf y}_i,{\bf y}_j$, we define the distance as
\begin{equation}
r_{ij}=d({\bf y}_i,{\bf y}_j)=\sum_{l=0}^{m-1}|x_{i+lp}-x_{j+lp}|.
\end{equation}
If the distance is less than a present number $r$, we say that these two vectors are correlated.
The correlation integral is defined as
\begin{equation}
C_m(r)=\frac{1}{N_m^2}\sum_{i,j=1}^{N_m}H(r-r_{ij})
\end{equation}
where $H$ is the Heaviside function
\begin{equation}
H(x)=\left\{\begin{array}{l}1,\quad {\rm if}\ x>0\\
0,\quad {\rm if}\ x\le 0 \end{array}\right.
\end{equation}
For a proper choice of $m$ and not too big a value of $r$, it has been 
shown by Grassberger and
Procaccia$^{\cite{GP2}}$ that the correlation integral $C_m(r)$ behaves like
\begin{equation}
C_m(r)\ \propto\ r^{D_2(m)}.
\end{equation}
Thus one can define correlation dimension as
\begin{equation}
D_2=\lim_{m\longrightarrow\infty}D_2(m)=\lim_{m\longrightarrow\infty}\lim_{r\longrightarrow 0}
\frac{\ln C_m(r)}{\ln r}.
\end{equation}
For more details of $D_2$, the reader can refer to ref.\cite{Hao89}. 

\ \ To deal with practical problem, one usually choose $p=1$. From 
Page 346 of ref.\cite{Hao89}, if
we choose an sequence $\{r_i:\ 1\le i\le n\}$ such that $r_1<r_2<r_3<\cdots <r_n$,
 then
in the $\ln r-\ln C_m(r)$ plane, we can look for a scaling region. Then the slop of the 
scaling region is $D_2(m)$. When $D_2(m)$ dose not change with $m$ increasing, we can take
this $D_2(m_0)$ as the estimate value of $D_2$.  We calculate the correlation dimension of
 some DNA sequences using the method introduced above. From the $\ln r-\ln C_m(r)$ figures of these sequences of different value of embedding
 dimension $m$, we find that it is suitable to choose $m=7$. For example, 
 we give the $\ln r-\ln C_m(r)$
 figure of Phage's 5'UTR sequence when $m=7,8$ (Figure \ref{correlfig}). 
 We take the region from the third point to the 17th point
 (from left to right) as the scaling region   

To study time series,
Hurst$^{\cite{hurst}}$ invented a new statistical method --- 
{\it the rescaled range analysis} ($R/S$ analysis), later on 
B.~B.~Mandelbrot$^{\cite{mandelbrot}}$ and J. Feder $^{\cite{feder}}$ transplanted 
$R/S$ analysis into fractal analysis.
For any time series  $x=\{x_k\}_{k=1}^N$ and 
any $2\le n\le N$, one can define

\be <x>_{n}=\frac{1}{n}\sum_{i=1}^{n}x_i \ee

\be X(i,n)=\sum_{u=1}^i[x_u-<x>_{n}] \ee

\be R(n)=\max_{1\le i\le n}X(i,n)-\min_{1\le i\le n}X(i,n) \ee

\be S(n)=[\frac{1}{n}\sum_{i=1}^{n}(x_i-<x>_{n})^2]^{1/2}. \ee

Hurst found that

\be R(n)/S(n) \ \sim\ (\frac{n}{2})^H. \ee

 $H$ is called {\it Hurst exponent}.

\begin{figure}
\centerline{\epsfsize=10cm \epsfbox{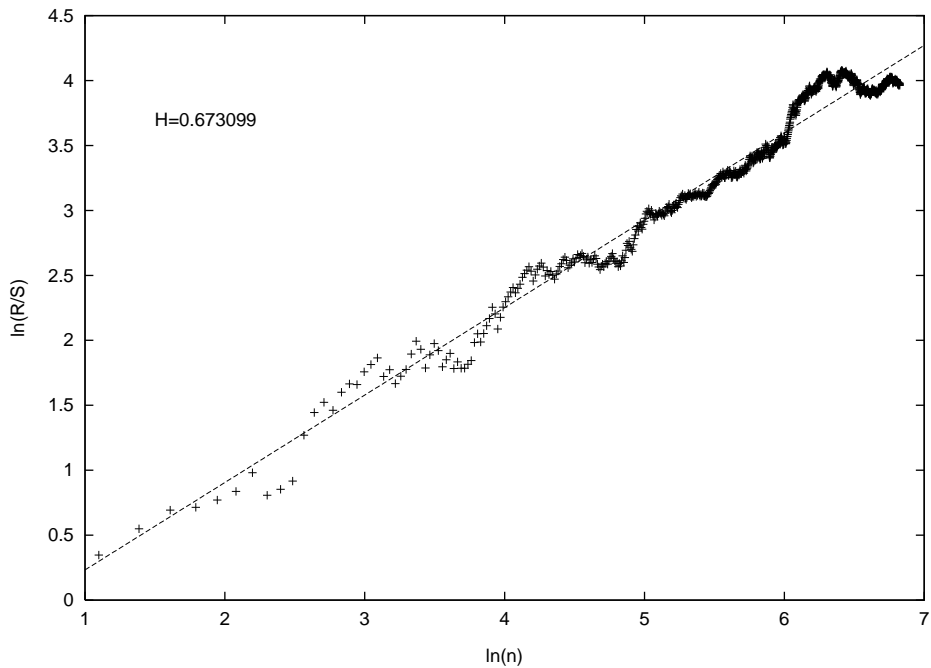}}
\caption{ An example of $R/S$ analysis of DNA sequence}
\label{rsfig}
\end{figure}

  As $n$ changes from 2 to $N$, we obtain $N-1$ points 
in $\ln(n)$ v.s. $\ln(R(n) / S(n))$
 plane. Then  we can calculate Hurst exponent $H$ of DNA sequence $s$ 
using the 
least-square linear fit. 
As an example, we plot the graph of $R/S$ analysis of an exon segment $s$ 
of mouse' DNA
sequence (bp 1730-- bp 2650 of the record with Accession AF033620 in Genbank) 
in Figure \ref{rsfig}.

\vskip 3mm

  The Hurst exponent is usually used as a measure of complexity. 
From Page 149 of Ref.[22], the trajectory 
of the record is a curve with a fractal dimension $D=2-H$. Hence 
a smaller $H$ means a more complex system.
When applied to fractional Brownian motion, if $H > 1/2$, 
the system is said to be {\em  persistent}, which means that if
for a given time period $t$, the motion is along one direction,
then in the succeeding $t$ time, it's more likely that the motion
will follow the same direction. While for system with $H < 1/2$,
the opposite holds, that is, {\em antipersistent}. But when $H=1/2$,
the system is Brown motion, and is random.

\section{ Data and results. }
\ \ \ \ More than 18  bacterial complete genomes are now available 
in public databases
. There are four Archaebacteria: Archaeoglobus fulgidus (aful), 
Pyrococcus horikoshii (pyro), Methanococcus jannaschii (mjan), 
and Methanobacterium thermoautotrophicum (mthe); four Gram-positive 
Eubacteria: Mycobacterium tuberculosis (mtub), Mycoplasma pneumoniae (mpneu),
 Mycoplasma genitalium (mgen), and Bacillus subtilis (bsub). The others 
 are Gram-negative Eubacteria:  one hyperthermophilic bacterium Aquifex 
 aeolicus (aquae); six proteobacteria: Rhizobium sp. NGR234 (pNGR234), 
 Escherichia coli (ecoli), Haemophilus influenzae (hinf),
  Helicobacter pylori J99 (hpyl99), Helicobacter pylori 26695
 (hpyl) and Rockettsia prowazekii (rpxx);  two 
 chlamydia Chlamydia trachomatis (ctra) and Chlamydia pneumoniae (cpneu), 
 and one cyanobacterium 
 Synechocystis PCC6803 (synecho).
 \begin{table}
\caption{Average of $D_2$ of genes of 18  bacteria.}
\begin{center}
\begin{tabular}{|c|l|l|}
\hline\hline
      Average of $D_2$ & species of Bacterium &Category\\
 \hline
 2.805 & Mycoplasma genitalium (mgen) & Gram-positive Eubacteria\\
 2.827 & Methanococcus jannaschii (mjan) & Archaebacteria\\
 2.872 & Rockettsia prowazekii (rpxx) & Proteobacteria\\
 2.883 & Helicobacter pylori 26695 (hpyl) & Proteobacteria\\
 2.908 & Helicobacter pylori J99 (hpyl99) & Proteobacteria\\
 2.938 & Haemophilus influenzae (hinf) & Proteobacteria\\
 2.940 & Mycoplasma pneumoniae (mpneu) & Gram-positive Eubacteria\\
 2.950 & Mycobacterium tuberculosis (mtub) & Gram-positive Eubacteria\\
 2.990 & Bacillus subtilis (bsub) & Gram-positive Eubacteria\\
 \hline\hline
 3.011 & Aquifex aeolicus (aquae) & hyperthermophilic bacterium\\
 3.012 & Pyrococcus horikoshii (pyro) & Archaebacteria\\
 3.013 & M. thermoautotrophicum& Archaebacteria (mthe)\\
 3.016 & Archaeoglobus fulgidus (aful) & Archaebacteria\\
 3.020 & Chlamydia trachomatis (ctra) & Chlamydia \\
 3.024 & Chlamydia pneumoniae (cpneu) & Chlamydia \\
 3.028 & Synechocystis PCC6803 (synecho) & Cyanobacteria \\
 3.047 &  Rhizobium sp. NGR234 (pNGR234) &Proteobacteria\\
 3.060 & Escherichia coli (ecoli) &Proteobacteria\\
 \hline\hline
 \end{tabular}
 \end{center}
 \label{homotab1}     
\end{table}

 For a given bacterium, we calculate the correlation dimension and Hurst exponent of
 each CDS sequence (i.e. the coding sequence) in its complete genome first (the 
 results is shown in Figure 3 and 
 Figure 4), then
 calculate the average of these two quantities. We find that the average of Hurst 
 exponents of 18 bacteria are almost equal (range being from 0.538 to 0.590).
 But the differences among the values of average of correlation dimensions of
  these bacteria are larger. One can see Table 1 (from top to bottom, the value of
  $D_2$ become larger).

\section{Discussion and conclusions}

\ \ \ \ Although the existence of the archaebacterial urkingdom has been accepted by many 
biologists, the classification of bacteria is still a matter of controversy$^{\cite{iwabe}}$.
The evolutionary relationship of the three primary kingdoms (i.e. archeabacteria, eubacteria
and eukaryote) is another crucial problem that remains unresolved$^{\cite{iwabe}}$. 

From Table 1, we can roughly divide bacteria into two class first, 
the average of $D_2$ of one class is less than 3.0, that of another class is greater
than 3.0.
We can see that the classification of bacteria using the average of 
$D_2$ is almost coincide with the original classification of bacteria. 
 Archaebacteria  gather with each other except mjan.
 Gram-positive bacteria get together except mgen. Chlamydia also gather
 with each other. Proteobacteria is divided into two sub-category: rpxx, hpyl, hpyl99 and
 hinf belong to one sub-category; pNGR234 and ecoli belong another sub-category.
 
  A surprising feature shown in Table 1 is that Aquifex aeolicus is 
linked closely with the Archaebacteria. We noticed that Aquifex, like most 
Archaebacteria, is hyperthermophilic. It has previously been shown that 
Aquifex has close relationship with Archaebacteria from the gene comparison of
 an enzyme needed for the synthesis of the amino acid 
 trytophan$^{\cite{pennisi}}$. 
 Our result, from the comparison of the complete genome, shows that the 
 case is even more worse. Such strong correlation on the level of complete 
 genome between Aquifex and Archaebacteria is not easily accounted for by lateral 
 transfer and other accidental events$^{\cite{pennisi}}$.

  We calculate the average of correlation dimensions and Hurst exponents of 
  genes in all 16 chromosome of Saccharomyces cerevisiae (yeast), they are 
  3.018 and 0.579 respectively. From Table 1, one can see that Archaebacteria
  and Chlamydia are linked more closely with yeast which belongs to eukaryote than 
  other category of bacteria. There are several reports (such as Ref. \cite{lhb}),
  in some RNA and protein species, archeabacteria are much more similar in sequences
  to eukaryotes than to eubacteria. Our present result supports this point of view. 
  
  We also randomly produce a sequence of length 3000 consisting of symbols from 
  the alphabet
$\{A,T,G,C\}$. The
correlation dimension is 1.02883. From Table \ref{homotab1}, Fig. 3,
 we can conclude that all CDS sequences are far from random sequences.
 Since the Hurst exponent of random sequence is 0.5. From Fig. 4, 
 we can  see that correlation dimension is well than Hurst exponent
  when we compare real
 DNA sequence with a random sequence on
the alphabet $\{A,T,G,C\}$.

   In Ref. \cite{YC}, we find the Hurst exponent is a good exponent to distinct
 different functional regions, but now we only consider the same kind of functional
 region (i.e. they are all genes), it is reasonable that the average of Hurst 
 exponent do not change much. Fortunately, now we can use the average of correlation
 dimension to distinguish different species.

\section*{ACKNOWLEDGMENTS}
\ \ \   The  authors would like to express their thanks to Prof. Bai-lin Hao for
 reading the manuscript carefully, encouragement and many good suggestions.

\pagebreak 
 \begin{figure}
\centerline{ \epsfbox{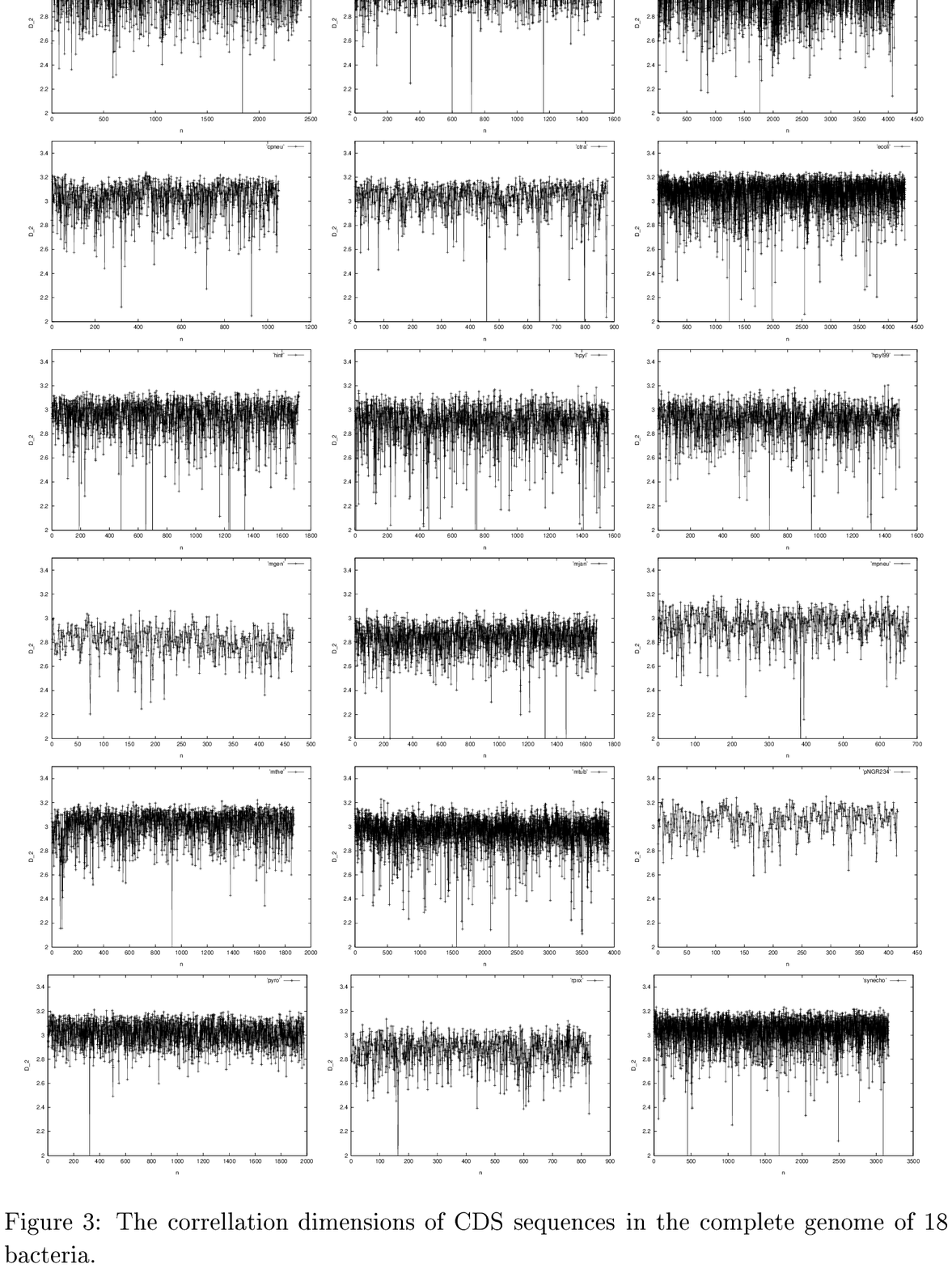}
  } 
%\label{fig3}
\end{figure}

\pagebreak
\begin{figure}
\centerline{ \epsfbox{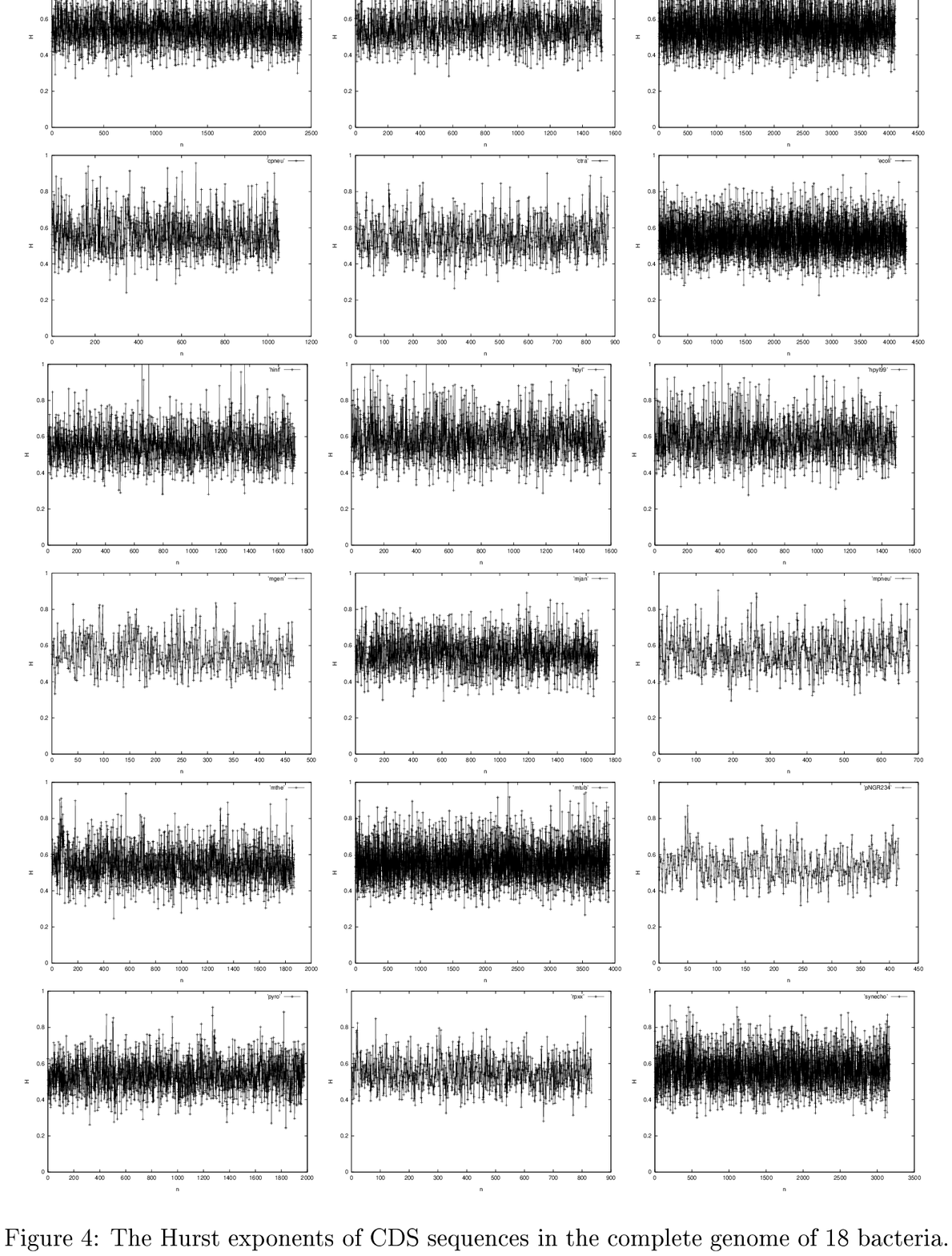}
 } 
%\label{fig4}
\end{figure}

\end{document}